\documentclass[showpacs,aps,graphicx,twocolumn]{revtex4}
\usepackage{amsmath}
\usepackage{amscd}
\usepackage{graphicx}
\begin{document}

\title{Efficient polarization entanglement concentration for electrons with charge detection\footnote{Published in
Phys. Lett. A 373, 1823 (2009)}}
\author{ Yu-Bo Sheng$^{a,b,c}$, Fu-Guo Deng$^{d}$\footnote{E-mail address: fgdeng@bnu.edu.cn}, Hong-Yu Zhou$^{a,b,c}$ }
\address{$^{a}$ The Key Laboratory of Beam Technology and Material
Modification of Ministry of Education, Beijing Normal University,
Beijing 100875, People's Republic of China\\
$^{b}$ Institute of Low Energy Nuclear Physics, and Department of
Material Science and Engineering, Beijing Normal University,
Beijing 100875, People's Republic of China\\
$^{c}$ Beijing Radiation Center, Beijing 100875, People's Republic
of China\\
$^{d}$ Department of Physics, Applied Optics Beijing Area Major
Laboratory, Beijing Normal University, Beijing 100875, China}
\date{\today }

\begin{abstract}
We present an entanglement concentration protocol for electrons
based on their spins and their charges. The combination of  an
electronic polarizing beam splitter and a charge detector
functions as a parity check device for two electrons, with which
the parties can reconstruct maximally entangled electron pairs
from those in a less-entanglement state nonlocally. This protocol
has a higher efficiency than those based on linear optics and it
does not require the parties to know accurately the information
about the less-entanglement state, which makes it more convenient
in a practical application of solid quantum computation and
communication.
\\
\textbf{ keywords:} quantum physics, entanglement concentration,
electrons,
 charge detection, quantum computation
\end{abstract}
\pacs{03.67.Mn, 03.67.Pp, 03.67.Hk, 42.50.Dv} \maketitle

Entanglement plays an important role in quantum information
processing and transmission \cite{computation1,computation2,RMP}.
For most of the practical quantum computation and communication
protocols, the maximally entangled states are usually required.
However, an entangled quantum system transmitted in a realistic
quantum channel (such as a fiber or a free space) will suffer from
noise, which will degrade the quality of its entanglement, and
then a maximally entangled state will become a mixed  one or a
less-entanglement one. Entanglement purification provides us an
essential way to increase the entanglement of quantum systems in a
mixed state. Several entanglement purification protocols have been
proposed \cite{C.H.Bennett1,Deutsch,Pan1,Simon,shengpra1} since
the first one based on controlled-not (CNOT) gates and bilateral
operations was proposed by Bennett \emph{at al.}
\cite{C.H.Bennett1} in 1996. For instance, Deutsch \emph{et al.}
\cite{Deutsch} improved the efficiency and decreased the
difficulty in the first one. Pan \emph{et al.} \cite{Pan1}
proposed an entanglement purification protocol (EPP) with
polarizing beam splitters (PBSs) and sophisticated single-photon
detectors in 2001. Simon and Pan \cite{Simon} presented an EPP for
a parametric down-conversion (PDC) source in 2003. Recently, we
\cite{shengpra1} introduced a scheme for polarization-entanglement
purification based on PDC sources with cross-Kerr nonlinearity,
which is not only suitable for an ideal entanglement source but
also for a PDC source.

Different from entanglement purification, entanglement
concentration is used to distill a set of less-entanglement pure
states for obtaining a subset of maximally entangled states. This
topic is interesting as the process for storing quantum systems
and even producing entangled states with asymmetrical devices
usually makes the entangled quantum systems become
less-entanglement ones. The first entanglement concentration
protocol (ECP), named Schmidt projection method,  was proposed by
Bennett \emph{et al.} \cite{C.H.Bennett2} in 1996. In their
protocol, they need some collective measurements, which are hard
to manipulate in experiment at present. They also need to know the
accurate information of the less-entanglement state. There is
another type of ECP, named entanglement swapping
\cite{swapping1,swapping2} in which collective Bell-state
measurements are required. Two similar ECPs were proposed
independently by Yamamoto \emph{et al.} \cite{Yamamoto} and Zhao
\emph{et al} \cite{zhao1}. In their protocol, they use some PBSs
to make a parity check for two photons. However, for
distinguishing the four-mode instances from others, both the
parties should possess some sophisticated single-photon detectors.
In 2008, we \cite{shengpra2} proposed an ECP based on the
cross-Kerr nonlinearities.  By iteration of this protocol, the
whole efficiency and the yield are higher than those with linear
optical elements. However, for getting a higher efficiency and
yield, a strong cross-Kerr media or an intense coherent beam is
required.

Currently, one one hand, most of the entanglement purification and
concentration protocols are focused on photons. On the other hand,
quantum communication and computation can also be achieved with
conduction electrons since Beenakker  \emph{et al.}
\cite{beenakker} broke through the obstacle of the no-go theorem
\cite{nogo} in 2004. An electron system has both its spin degree
of freedom and its charge degree of freedom. Moreover, Spin and
charge commute, so a measurement of the charge leaves the spin
qubit unaffected \cite{beenakker}. With the charge detector
\cite{cd} which can distinguish the occupation number one from the
occupation number 0 and 2, but cannot distinguish between 0 and 2,
people can construct CNOT gates \cite{beenakker} and charge qubits
\cite{charge}, entangle spins \cite{paritybox}, and prepare
cluster states and a multipartite entanglement analyzer
\cite{cluster}.

In this Letter, we present an electronic entanglement
concentration protocol with the aid of charge detection, following
some ideas in Schmidt projection method \cite{C.H.Bennett2} and
quantum erasure \cite{erasure}. The combination of a polarizing
beam splitter (PBS) and a charge detector functions as a parity
check device for electron spins with nondestructive measurements,
with which the parties can reconstruct maximally entangled
electron pairs from those in a less-entanglement pure state
efficiently. This protocol does not require the parties to know
accurately the information about the less-entanglement states,
i.e., their coefficients. Compared with the protocols based on
linear optical elements, it has a higher efficiency as the states
in unsuccessful instances in the first entanglement concentration
process can also be concentrated probabilistically in the next
round.

Now, we detail how our ECP works. Let us consider two pairs of
entangled electrons  in the following unknown polarization states:
\begin{eqnarray}
|\Phi\rangle_{a_1b_1} &=&
\alpha|\uparrow\rangle_{a_1}|\uparrow\rangle_{b_1}
+\beta|\downarrow\rangle_{a_1}|\downarrow\rangle_{b_1},\nonumber\\
|\Phi\rangle_{a_2b_2} &=&
\alpha|\uparrow\rangle_{a_2}|\uparrow\rangle_{b_2}
+\beta|\downarrow\rangle_{a_2}|\downarrow\rangle_{b_2},\label{eq1}
\end{eqnarray}
where  $|\uparrow\rangle$ and $|\uparrow\rangle$  are the spin up
state and the spin down state, respectively, and
$|\alpha|^{2}+|\beta|^{2}=1$. Alice owns the electrons $a_{1}$ and
$a_{2}$, and Bob owns the electrons $b_{1}$ and $b_{2}$. Here we
call the mode  $a_{1}b_{1}$ the upper mode, and
 $a_{2}b_{2}$ the lower mode, shown in Fig.1. The original nonmaximally entangled
state of the four electrons can be written as:
\begin{eqnarray}
|\Psi\rangle &\equiv& |\Phi\rangle_{a_1b_1}\otimes
|\Phi\rangle_{a_2b_2}=\alpha^{2}|\uparrow\rangle_{a_1}|
\uparrow\rangle_{b_1}|\uparrow\rangle_{a_2}|\uparrow\rangle_{b_2}\nonumber\\
&+& \alpha\beta|\uparrow\rangle_{a_1}|\uparrow\rangle_{b_1}|\downarrow\rangle_{a_2}|\downarrow\rangle_{b_2}
+\alpha\beta|\downarrow\rangle_{a_1}\downarrow\rangle_{b_1}|\uparrow\rangle_{a_2}|\uparrow\rangle_{b_2} \nonumber\\
&+&
\beta^{2}|\downarrow\rangle_{a_1}|\downarrow\rangle_{b_1}|\downarrow\rangle_{a_2}|\downarrow\rangle_{b_2}.\label{eq2}
\end{eqnarray}
Before the two electrons $a_{2}$ and $b_{2}$ are transmitted to
Alice and Bob, respectively, two electronic half-wave plates are
used to transfer $|\uparrow\rangle$ to $|\downarrow\rangle$ or
vice versa. The whole state of the two electron pairs becomes:
\begin{eqnarray}
|\Psi\rangle^{'} &=&\alpha^{2}|\uparrow\rangle_{a_1}|\downarrow\rangle_{a_3}|\uparrow\rangle_{b_1}|
\downarrow\rangle_{b_3}
+ \alpha\beta|\uparrow\rangle_{a_1}|\uparrow\rangle_{a_3}|\uparrow\rangle_{b_1}|\uparrow\rangle_{b_3}\nonumber\\
&+&
\alpha\beta|\downarrow\rangle_{a_1}|\downarrow\rangle_{a_3}|\downarrow\rangle_{b_1}|\downarrow\rangle_{b_3}%
+ \beta^{2}|\downarrow\rangle_{a_1}|\uparrow\rangle_{a_3}|
\downarrow\rangle_{b_1}|\uparrow\rangle_{b_3}.
\label{staterotation}
\end{eqnarray}
We use $a_{3}(b_{3})$ to substitute  $a_{2}(b_{2})$ after the
$90^{\circ}$ rotation $R_{90}$ which transfers $|\uparrow\rangle$
to $|\downarrow\rangle$ and vice versa. It is obvious that the two
terms
$|\uparrow\rangle_{a_1}|\uparrow\rangle_{a_3}|\uparrow\rangle_{b_1}|\uparrow\rangle_{b_3}$
and
$|\downarrow\rangle_{a_1}|\downarrow\rangle_{a_3}|\downarrow\rangle_{b_1}|\downarrow\rangle_{b_3}$
have the same coefficient of $\alpha\beta$, and the other two
terms are different ($\alpha^{2}$ or $\beta^{2}$). Bob lets the
two electrons $b_{1}$ and $b_{3}$ pass through a PBS which fully
transmits $|\uparrow\rangle$ polarization electrons and fully
reflects $|\downarrow\rangle$ electrons.

After the PBS, one can see that in Bob's laboratory, the states
$|\uparrow\rangle_{b_1}|\uparrow\rangle_{b_3}$ and
$|\downarrow\rangle_{b_1}|\downarrow\rangle_{b_3}$ will make each
of the two spatial modes $c_1$ and $c_2$ contains one electron.
The charge detector $P$ will detect only one electron with a
nondestructive measurement. However, the state
$|\uparrow\rangle_{b_1}|\downarrow\rangle_{b_3}$ will make two
electrons in the lower mode $c_2$ and
$|\downarrow\rangle_{b_1}|\uparrow\rangle_{b_3}$ will make the two
electrons in the upper mode $c_1$, respectively,  which means that
the charge detector will detect two electrons or no electrons.

\begin{figure}[!h]
\begin{center}
\includegraphics[width=8cm,angle=0]{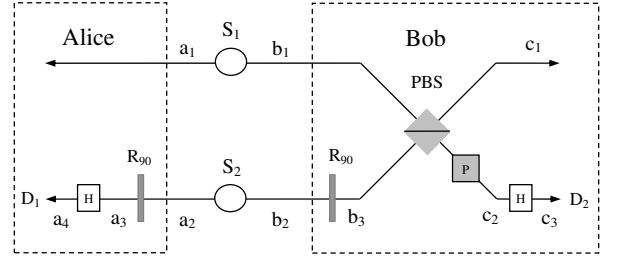}
\caption{Schematic diagram showing the principle of the proposed
entanglement concentration protocol for entangled electron pairs
with charge detection. Alice and Bob receive two pairs of identical
less-entanglement electrons which are sent from the two sources
$S_{1}$ and $S_{2}$, respectively. PBS: polarizing beam splitter. P
and H represent a charge detector and a Hadamard operation,
respectively. $R_{90}$ is an electronic half-wave plate which
transfers $\vert \uparrow\rangle$ to $\vert \downarrow\rangle$ and
vice versa.  The combination of PBS and P can make a parity check
for the spins of two electrons.}
\end{center}
\end{figure}

If the charge detector detects only one electron, the four
electrons will collapse to the state as:
\begin{eqnarray}
|\Psi\rangle^{''} &=&
\frac{1}{\sqrt{2}}(|\uparrow\rangle_{a_1}|\uparrow\rangle_{a_3}|
\uparrow\rangle_{c_1}|\uparrow\rangle_{c_2} \nonumber\\
&& \;\;\;\; +
|\downarrow\rangle_{a_1}|\downarrow\rangle_{a_3}|\downarrow\rangle_{c_1}|\downarrow\rangle_{c_2}).
\label{maxstate}
\end{eqnarray}
The probability for the charge detector to detect one electron in
each mode is $2|\alpha\beta|^{2}$.

The state described with Eq.(\ref{maxstate}) is a maximally
entangled state for four electrons.  It is easy to get a maximally
entangled two-electron state. We only need to perform a Hadamard
operation on each of the two electrons $a_{3}$ and $c_{2}$, and
then measure them with the basis $Z=\{\vert\uparrow\rangle, \vert
\downarrow\rangle \}$, shown in Fig.1. In detail, after the two
Hadamard (H) operations on  $a_{3}$ and $c_{2}$ (the H operation
completes the transformations $\vert \uparrow\rangle$
$\rightarrow$ $(\vert \uparrow\rangle + \vert
\downarrow\rangle)/\sqrt{2}$ and  $\vert \downarrow\rangle$
$\rightarrow$ $(\vert \uparrow\rangle - \vert
\downarrow\rangle)/\sqrt{2}$)), the whole state of the four
electrons becomes:
\begin{eqnarray}
 |\Psi\rangle^{'''}&=&\frac{1}{2\sqrt{2}}(|\uparrow\rangle_{a_1}|\uparrow\rangle_{c_1}
 +|\downarrow\rangle_{a_1}|\downarrow\rangle_{c_1})
 (|\uparrow\rangle_{a_4}|\uparrow\rangle_{c_3}\nonumber\\
&+&
|\downarrow\rangle_{a_4}|\downarrow\rangle_{c_3})+\frac{1}{2\sqrt{2}}(|\uparrow\rangle_{a_1}|\uparrow\rangle_{c_1}-
|\downarrow\rangle_{a_1}|\downarrow\rangle_{c_1})\nonumber\\
&&
(|\uparrow\rangle_{a_4}|\downarrow\rangle_{c_3}+|\downarrow\rangle_{a_4}|\uparrow\rangle_{c_3}).\label{statedistinguish}
\end{eqnarray}
The last step is to measure the spins of the electrons $a_4$ and
$c_3$ with the basis $Z$. If the two detectors $D_{1}$ and $D_{2}$
have the same results, the electron pair $a_{1}c_{1}$ will
collapse to the state:
\begin{eqnarray}
|\phi^{+}\rangle_{a_1c_1}=\frac{1}{\sqrt{2}}(|\uparrow\rangle_{a_1}|\uparrow\rangle_{c_1}
+|\downarrow\rangle_{a_1}|\downarrow\rangle_{c_1}).
\end{eqnarray}
Otherwise, we will get
\begin{eqnarray}
|\phi^{-}\rangle_{a_1c_1}=\frac{1}{\sqrt{2}}(|\uparrow\rangle_{a_1}|\uparrow\rangle_{c_1}
-|\downarrow\rangle_{a_1}|\downarrow\rangle_{c_1}).
\end{eqnarray}
Alice or Bob performs a phase-flip operation on her or his
electron to get $|\phi^{+}\rangle_{a_1c_1}$ . That is, a maximally
entangled two-electron state $|\phi^{+}\rangle$ can be generated
with the steps described above.

\begin{figure}[!h]
\begin{center}
\includegraphics[width=8cm,angle=0]{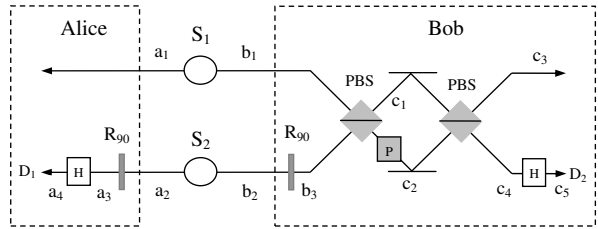}
\caption{The principle for obtaining some less-entanglement states
from the fail instances. Another PBS is used to make each spatial
mode contain only one electron. }
\end{center}
\end{figure}

In our protocol, the  charge detector $P$ is used to detect the
parity of two electrons. If the two electrons are in an even
parity (both spin up or down), the charge detector will detect
only one electron, and the less-entanglement state $\vert
\Phi\rangle$ can be concentrated to the maximally entangled state.
Otherwise, the charge detector will detect  0 or 2 electrons,
which means the entanglement concentration process fails. In this
time, the four-electron system collapses to another
less-entanglement state (without being normalized):
\begin{eqnarray}
|\Phi_1\rangle &=& \alpha^2
|\uparrow\rangle_{a_1}|\downarrow\rangle_{a_3}|\uparrow\rangle_{c_1}|\downarrow\rangle_{c_2}\nonumber\\
&+& \beta^2
|\downarrow\rangle_{a_1}|\uparrow\rangle_{a_3}|\downarrow\rangle_{c_1}|\uparrow\rangle_{c_2}.
\label{lessstate2}
\end{eqnarray}
Different from the ECPs with linear optics \cite{Yamamoto,zhao1},
the states in the unsuccessful instances in this protocol can also
be concentrated to the maximally entangled state in the next
round. We show the principle in Fig.2. Another PBS is added to
divide the two electrons into two different spatial modes $c_3$
and $c_4$. After the measurements on the electrons $a_4$ and
$c_5$, Eq.(\ref{lessstate2}) become:
\begin{eqnarray}
|\Phi_1\rangle^{'} = \alpha^2
|\uparrow\rangle_{a_1}|\uparrow\rangle_{c_3} \pm& \beta^2
|\downarrow\rangle_{a_1}|\downarrow\rangle_{c_3}.\label{eq9}
\end{eqnarray}
$'+'$ or $'-'$ depend on the facts that the results of $D_{1}$ and
$D_{2}$ are the same one or different ones, respectively. It is
obvious that Eq.(\ref{eq9}) has the same form as Eq.(\ref{eq1}).
We can pick up two pairs of electrons in the less-entanglement
state shown in Eq.(\ref{eq9}) and perform the similar
concentration process as our ECP does not require the two parties
to know accurately the information about the coefficients $\alpha$
and $\beta$. With the iteration of the entanglement concentration
process above, the efficiency of our protocol is higher than the
protocols based on linear optics \cite{Yamamoto,zhao1}.

It is straightforward to generalize this entanglement
concentration protocol to the case for multipartite pure entangled
states. Let us suppose that the pure entanglement states of $n$-
electron quantum  systems are
\begin{eqnarray}
|\Phi_2\rangle = \alpha \vert u\rangle
|\uparrow\rangle_{a_1}|\uparrow\rangle_{b_1} + \beta \vert
\overline{u}\rangle
|\downarrow\rangle_{a_1}|\downarrow\rangle_{b_1},
\end{eqnarray}
where $\vert u\rangle$ and $\vert \overline{u}\rangle$ are states
of $n-2$ electrons, which are owned by the other $n-2$ parties
(not Alice and Bob). The $n$ parties can accomplish the
entanglement concentration as the same as that discussed above by
replacing  $\alpha$ and $\beta$ in Eq.(\ref{eq1}) with $\alpha
\vert u\rangle$ and $\beta \vert \overline{u}\rangle$,
respectively. With the similar operations, the $n$ parties can
reconstruct maximally entangled $n$-electron
Greenberger-Horne-Zeilinger (GHZ) states
$|\phi^+_2\rangle_n=\frac{1}{\sqrt{2}}(\vert
u\rangle|\uparrow\rangle_{a_1}|\uparrow\rangle_{b_1} +  \vert
\overline{u}\rangle
|\downarrow\rangle_{a_1}|\downarrow\rangle_{b_1})$ from partially
entangled GHZ-class states $|\Phi_2\rangle$.

In conclusion, we propose an electronic entanglement concentration
protocol based on charge detection. We exploit the combination of
an electronic polarizing beam splitter and a charge detector to
distinguish the parity of  two electrons. Compared with other
ECPs, this protocol is simpler and convenient  as it does not
require collective measurements and sophisticated detectors.
Moreover, it does not require the parties to know accurately the
information about the less-entanglement state. By iterating the
entanglement concentration processes, this protocol has a higher
efficiency than those based on linear optics. These advantages
make our scheme have a good application in solid quantum
computation and communication.

This work is supported by the National Natural Science Foundation
of China under Grant No. 10604008, A Foundation for the Author of
National Excellent Doctoral Dissertation of China under Grant No.
200723, and  Beijing Natural Science Foundation under Grant No.
1082008.

\end{document}